\documentclass[pre,twocolumn,showpacs]{revtex4}

\usepackage{graphicx,amsmath}
\def\gs{\gtrsim}
\def\ls{\lesssim}
\def\be{\begin{equation}}
\def\en{\end{equation}}                  
\def\p{\partial} 
\newcommand{\bi}[1]{\mbox{\boldmath$#1$}}
\newcommand{\av}[1]{\langle{#1}\rangle}

\def\bea{\begin{equation}\begin{array}{rcl}}
\def\ena{\end{array}\end{equation}}

\def\asigma{\stackrel{\leftrightarrow}{\sigma}}
\def\aPi{\stackrel{\leftrightarrow}{\Pi}} 
\def\aw{\stackrel{\leftrightarrow}{W}}

\def\gdot{\dot{\gamma}}

\def\q{{\footnotesize{\it q}}\kern -5pt {\footnotesize{\it q}}}
\def\k{{\footnotesize{\it k}}\kern -5pt {\footnotesize{\it k}}}
\def\seq{\sim \kern -12pt \lower 5pt \hbox{$\displaystyle =$}}
\def\nnabla{\nabla\kern-3.3mm\nabla}

\def\ge{> \kern -12pt \lower 5pt \hbox{$\displaystyle =$}}
\def\le{< \kern -12pt \lower 5pt \hbox{$\displaystyle =$}}
\def\gs{> \kern -12pt \lower 5pt \hbox{$\displaystyle{\sim}$}}
\def\ls{< \kern -12pt \lower 5pt \hbox{$\displaystyle{\sim}$}}

\def\be{\begin{equation}}
\def\bea{\begin{eqnarray}}
\def\en{\end{equation}}
\def\ena{\end{eqnarray}}

\def\asigma{\stackrel{\leftrightarrow}{\sigma}}
\def\aPi{\stackrel{\leftrightarrow}{\Pi}} 
\def\aw{\stackrel{\leftrightarrow}{W}}

\def\q{\scriptsize q \keren -12pt \scriptsize q}

\def\p{\partial }

\begin{document}
\preprint{}
\title{Viscoelastic Phase Separation in Shear Flow}
\author{Tatsuhiro Imaeda$^1$, 
 Akira Furukawa$^2$, and Akira  Onuki$^2$}
\affiliation{$^1$Aichi Gakusen University, 
Toyota 471-8532,Japan\\
$^2$Department of Physics, Kyoto University, Kyoto 606-8502,Japan}
\date{\today}

\begin{abstract} 
We numerically investigate viscoelastic  phase separation in polymer solutions 
under shear using a time-dependent Ginzburg-Landau model. 
The gross variables in our model are the polymer volume fraction 
and a conformation tensor. 
The latter represents chain deformations and relaxes slowly 
on the rheological time giving rise to a large viscoelastic stress.  
The polymer and the solvent obey two-fluid dynamics 
in which the viscoelastic  stress acts asymmetrically on the polymer and, 
as a result, the stress and the diffusion are dynamically coupled.
Below the coexistence curve, interfaces appear 
with increasing the quench depth and the solvent regions act as a lubricant. 
In these cases the composition heterogeneity 
causes more enhanced viscoelastic heterogeneity and the macroscopic stress 
is decreased at fixed applied shear rate. 
We find  steady  two-phase states composed of the polymer-rich 
and solvent-rich regions, 
where the characteristic domain size is inversely proportional to the 
average shear stress for various shear rates.  
The deviatoric stress components exhibit large temporal fluctuations. 
The normal stress difference can take negative values transiently 
at weak shear.   
\end{abstract}

\maketitle

\section{Introduction}

In phase-separating polymer systems, 
domain morphologies are influenced by a number of factors, 
including the molecular weights, the composition, 
closeness to the critical point, and viscoelasticity \cite{Utracki,Onukibook}. 
Particularly when the two components have distinctly different 
viscoelastic properties as in semidilute polymer solutions and  
polymer blends of long and short chains,
unique interplays emerge between viscoelasticity and thermodynamic instability.
In such asymmetric systems, 
salient effects observed in experiments are as follows.   
First, in early stage spinodal decomposition,  
the kinetic coefficient $L(q)$ depends on the wave number $q$ as 
$L(q)/L(0) \sim (\xi_{\rm ve}q)^{-2}$ for $q$ larger  than the inverse of 
a viscoelastic length $\xi_{\rm ve}$ \cite{Scwahn92,Hashimoto2000}, 
where $\xi_{\rm ve}$ can be much longer than the gyration radius  
\cite{Doi-Onuki}. 
Second, in late stage spinodal decomposition, 
sponge-like network structures appear transiently in the asymmetric case  
\cite{HTanaka93and94,HTanaka96}. 
The physical origin of these effects  
is now ascribed to the stress-diffusion coupling in viscoelastic 
binary mixtures \cite{Doi-Onuki}. 
This coupling is predicted to give rise to various effects including 
non-exponential decay in dynamic light scattering \cite{Onukibook,Doi-Onuki} 
and shear-induced fluctuation enhancement. 
In the case of spinodal decomposition, 
the consequences of the dynamic coupling were studied analytically 
in the linear growth regime \cite{OnukiJCP} 
and numerically in the late stage coarsening 
\cite{Taniguchi,Tanaka-Araki,Yang}.

Flow effects on  phase separation in polymeric systems are even more dramatic  
\cite{Utracki,Larson1,Onukireview}. 
In this paper we consider a simple shear flow with mean velocity profile  
\be 
\av{{\bi v}}= \gdot y {\bi e}_x, 
\label{eq:1.2}
\en 
where the flow is in the $x$ direction, 
${\bi e}_x$ being the unit vector along the $x$ axis, 
and the mean velocity gradient $\gdot$ is in  the $y$ direction. 
Application of shear or extensional flow to 
viscoelastic systems in one-phase states sometimes  
induces a strong increase of the turbidity, indicating shear-induced 
composition heterogeneities or demixing. 
This is in marked contrast to shear-induced homogenization or mixing 
observed in near-critical fluids \cite{Onuki1,Beysens} 
and in ternary polymer mixtures \cite{Hashimoto_ternary,StringPRL}. 
In systems exhibiting shear-induced mixing, the entanglement 
effects are not severe and 
the hydrodynamic interaction is suppressed by shear 
\cite{Onukibook,Onukireview}. 
In semidilute polymer solutions near the coexistence curve with 
high molecular weights ($M\gs 2\times 10^6$), 
recent scattering experiments have most unambiguously 
detected shear-induced demixing 
\cite{Wu,Hashi1,Hashimoto97,Saito,Egmond,Han,Boue,Pine2}. 
Rheological effects in sheared polymer solutions are also conspicuous, 
which include large stress fluctuations upon demixing by shear 
\cite{Lodge,Peterlin}   
and  a second overshoot 
in the shear stress 
as a function of time after application of shear 
\cite{Hashimoto97,Larson_Macro}.    
Here microscope pictures 
of composition heterogeneities  in polymer solutions 
and asymmetric polymer blends under shear 
are informative \cite{Moses,Hobbie}.

Theoretically, for sheared polymer solutions,  
relevance of the dynamical coupling was first pointed out  by 
Helfand and Fredrickson \cite{Helfand}.
Some linear calculations for small fluctuations were also performed 
in Ginzburg-Landau schemes, where a conformation tensor represents 
the chain deformations \cite{OnukiPRL,Milner93,Ji,Fredrickson2002}. 
Numerical analysis (still in two dimensions)  
using such schemes gave insights into the nonlinear shear effects 
\cite{Onukireview,Onuki97,Okuzonop,Yuan,Yuan-Kawakatsu}.
That is, slightly above the coexistence curve, 
composition heterogeneities $\delta\phi$ on mesoscopic spatial scales 
emerge  with increasing $\gdot$.  
The amplitude of $\delta\phi$  can even be 
of the order of the average $\av{\phi}$, while there are  no clear interfaces. 
If this takes place, the system becomes turbid resulting in shear-induced 
{\it phase separation} or {\it demixing} observed above the coexistence curve.
We remark that similar Ginzburg-Landau models \cite{Ol,Ra} 
have been used to analyze shear-banding effects in wormlike micellar systems 
 \cite{rheo_chaos1,rheo_chaos2,Yuan_band}.

In entangled polymers the rheological relaxation time $\tau$ 
can be very long \cite{PGbook}, 
so experiments in the Newtonian and non-Newtonian regimes are both possible, 
where the Deborach number $De =\gdot \tau$ is larger or smaller than 1, 
respectively.  
In semidilute solutions, the strong composition-dependence of the 
solution viscosity $\eta =\eta (\phi)$ can lead to shear-induced 
fluctuation enhancement for weak shear $\gdot\tau \ll 1 $ \cite{Helfand},  
while the normal stress effect comes into play in the non-Newtonian regime 
 $\gdot\tau \gs  1 $ \cite{Onukibook,Saito,OnukiPRL}.

Without viscoelasticity a number of groups have performed simulations 
of phase separation in sheared simple fluids \cite{Ohta,Ber}. 
However, simulations have been still rare for  sheared viscoelastic fluids 
\cite{Onukireview,Onuki97,Okuzonop,Yuan,Yuan-Kawakatsu}.  
In this paper, we will examine nonlinear dynamic regimes  of sheared polymer 
solutions in theta solvent below the coexistence curve.  
Both in simple and polymeric fluids, if the system is quenched below 
the coexistence curve under shear, the domain growth is eventually stopped 
 and dynamically steady states are realized. 
For Newtonian fluids dynamics 
in sheared two-phase states have long been studied  
\cite{Onukibook,StringPRL,Fri,Migler}, 
but for viscoelastic fluids such nonequilibrium 
effects remain almost unexplored.

The organization of this paper is as follows. 
In Sec. II, we will explain our theoretical scheme.  
In Sec. III, we will present our numerical results in 
two dimensions below the coexistence curve 
for various shear rates and polymer volume fractions. 
Section IV summarizes new results and gives some predictions. 
\setcounter{equation}{0}

\section{Theoretical Background} 
In this section we briefly survey our theoretical framework 
to discuss viscoelastic phase separation under the shear flow (\ref{eq:1.2}). 
The gross variables in our model are the polymer volume fraction $\phi$
and a conformation tensor $\aw$. 
The latter represents chain deformations and relaxes slowly  
on the rheological time giving rise to a large viscoelastic stress. 
In Sec. II A we first present the viscoelastic Gintzburg-Landau free energy, 
and then in Sec. II B the dynamic model is given.  
The rheological properties of our dynamic equations are also discussed 
in Sec. II C.
See Refs. \cite{Onukibook,Onukireview} for
the details of our theoretical framework.

\subsection{Viscoelastic Ginzburg-Landau Free Energy}

Phase behavior of polymer solutions near the coexistence 
curve is usually described in term of the Flory-Huggins free energy density 
for the polymer volume fraction $\phi$ assumed to be much smaller than 1 
\cite{PGbook}, 
\be
f_{\rm FH} = \frac{k_{\rm B} T}{v_0} {\bigg [}
{\frac{\phi}{N}} \ln\phi + ({\frac{1}{2}}-\chi)\phi^2 
+{\frac{1}{6}}\phi^3 {\bigg ]},
\label{eq:2.1}
\end{equation}
where $v_0$ is the volume of a monomer 
($=a^3$ with $a$ being the monomer size in three dimensions) and 
$\chi$ is the interaction parameter dependent on the temperature $T$ 
(being equal to $1/2$ at the theta condition). 
At the critical point we have $\phi= \phi_{\rm c}=N^{-1/2}$ 
and $N^{1/2} (1-2\chi)=-2$. 
In the following it is convenient to scale $\phi$ and $2\chi-1$ as  
\be 
\Phi=\phi/\phi_{\rm c},\quad u=  N^{1/2} (2\chi-1). \label{eq:2.2}
\en 
In Fig.1 we show the phase diagram in 
the plane of $\phi/\phi_{\rm c}$ and $N^{1/2} (1-2\chi)=-u$.
The spinodal curve in Fig.1 is obtained from 
$(\p^2 f_{\rm FH}/\p \phi^2)_T=0$ and is written as $u=\Phi+\Phi^{-1}$.

To describe the viscoelastic effects on the composition inhomogeneities, 
it is convenient to introduce a tensor dynamic variable $\aw={\{}W_{ij}{\}}$, 
 which is a symmetric tensor representing chain conformations 
undergoing deformations \cite{OnukiPRL,Milner93,Ji,Grmela,Beris}.  
As shown in the following, 
the deviation $\delta W_{ij}= W_{ij}- \delta_{ij}$ 
gives rises to a network stress.

The Ginzburg-Landau free energy functional 
due to the fluctuations of $\phi$ and $\aw$ is written as
\be
F= \int  d{\bi r}{\bigg [}f_{\rm FH}+{\frac{1}{2}}C|\nabla \phi|^2
+{\frac {1}{4}}G Q(\aw){\bigg ]}. \label{eq:2.8}
\en 
The coefficient $C$ of the gradient term is calculated in the random 
 phase approximation \cite{PGbook} in the semidilute regime as 
\be
C = (k_{\rm B} T/18v_0)a^2/\phi.  
\label{eq:2.9}
\en 
We shall see that $G$ has the meaning of the shear modulus for 
small deformations changing rapidly compared with the rheological 
relaxation time $\tau$.  
It is assumed to be of the scaling form,   
\be
G = (k_{\rm B} T/v_0)g \phi^{\alpha}, \label{eq:2.11}
\en
where $g$ is an important  dimensionless parameter 
in our simulations, and is order of 1 in theta solvent
\cite{Onukibook,Onukireview}.  
Although experiments indicated $\alpha \cong 2.25$ 
\cite{Adam,Noda}, we will set $\alpha =3$ for simplicity. 
The simplest form for $Q(\aw)$ is given by 
\be
Q(\aw)= \sum_{ij} (\delta W_{ij})^2. \label{eq:2.13}
\en 

\subsection{Dynamic Equations}

We next present the dynamic equations assuming the two-fluid dynamics 
for the polymer and the solvent with the velocities, 
${\bi v}_{\rm p} $ and ${\bi v}_{\rm s} $, respectively, 
including the new variable $\aw$ \cite{OnukiPRL,Milner93,Ji,Ol}. 
These equations may be treated as Langevin equations 
with thermal noise terms, but we will neglect the noise terms hereafter.    
Formal frameworks for viscoelastic 
fluids have also been discussed in the literature \cite{Grmela,Beris}.

We assume that the mass densities of pure polymer and solvent are the same.  
Then the polymer mass fraction $\rho_{\rm p} /\rho$ coincides with $\phi$. 
It obeys 
\bea
{\frac{\partial}{ \partial t}}\phi &=& -\nabla\cdot (\phi {\bi v}_{\rm p} ) 
\nonumber\\ 
&=& -\nabla \cdot (\phi{\bi v})-\nabla\cdot[\phi(1-\phi){\bi w}], 
\label{eq:2.14}
\ena 
where 
\be
{\bi w}= {\bi v}_{\rm p} - {\bi v}_{\rm s}  
\label{eq:2.15}
\en 
is the relative velocity between polymer and solvent.
The mean  velocity $\bi v$ is defined by
\be
{\bi v}= \phi {\bi v}_{\rm p} + (1-\phi){\bi v}_{\rm s} .
\label{eq:2.16}
\en 
The two velocities ${\bi v}_{\rm p} $ and  ${\bi v}_{\rm s} $  
are expressed as  
${\bi v}_{\rm p} ={\bi v}+(1-\phi){\bi w}$ and 
${\bi v}_{\rm s} ={\bi v}-\phi{\bi w}.$ 
For simplicity, we assume the incompressibility condition 
for the average  velocity, 
\be 
\nabla\cdot{\bi v}=0.  
\label{eq:2.18}
\en 
On the other hand, $\nabla\cdot{\bi w}$ is 
nonvanishing in general and  gives rise to diffusion 
in (\ref{eq:2.14}) for small deviations around equilibrium.

Because $\aw$ represents the network deformation, its motion is determined 
by the polymer velocity ${\bi v}_{\rm p} $ and 
its simplest dynamic equation is of the form,
\bea
\bigg [{\frac{\partial }{\partial t}}
 &+& {\bi v}_{\rm p} \cdot\nabla \bigg ] W_{ij}
- \sum_{k}( D_{ik}W_{kj} + W_{ik}D_{jk} ) \nonumber\\
&=& -{\frac{1}{\tau^*}}\delta W_{ij},
\label{eq:2.19}
\ena
where $D_{ij}= {\partial} v_{{\rm p} i}/{\partial x_{j}}$ 
 is the  gradient of the polymer velocity. 
The left hand side of (\ref{eq:2.19}) is called the upper convective 
time derivative in the rheological literature \cite{Larsonbook}, 
which keeps the frame invariance of the tensor properties of $W_{ij}$.  
In (\ref{eq:2.41}) below, we will assume that 
the relaxation time  $\tau^*$ on the right hand side  
depends on ${\delta}{W_{ij}}$ as well as $\phi$. 
The usual rheological 
relaxation time $\tau$ is  obtained in  the Newtonian limit:
\be 
\tau = \lim_{{\delta}{\aw} \rightarrow 0}\tau^* .
\label{eq:2.21}
\en 
In the problem of shear-banding flow,  
some authors replaced $1/\tau^*$ in (\ref{eq:2.19}) 
by   $(1/\tau^*)(1- \ell^2\nabla^2)$ \cite{Ol,Yuan_band}.

The total stress tensor $\aPi=\{\Pi_{ij}\}$ is expressed as 
\cite{Onukibook,Onukireview} 
\bea
\Pi_{ij} &=& p \delta_{ij} 
+ C (\nabla_i\phi)(\nabla_j\phi)-
\sigma_{{\rm p} {ij}} \nonumber\\
 &-&\eta_0( \nabla_i v_j+\nabla_j v_i ),\label{eq:2.22}
\ena 
where $p$ is a pressure, 
$\nabla_i= \p/\p x_i$, 
${\asigma}_{\rm p} = \{ {\sigma}_{{\rm p}  ij} \}$ 
is the network stress arising from 
the deviation $\delta W_{ij}$, 
and the last term is the viscous stress 
tensor with  $\eta_0$ being 
the solvent viscosity. Assuming low Reynolds number flows 
and  setting $\p {\bi v}/\p t={\bi 0}$ (the Stokes 
approximation), we obtain  \cite{Onukireview} 
\be 
\nabla\cdot\aPi= -\nabla p_1  +{\bi {\cal F}}_{\rm p}     +   
 \eta_0 \nabla^2 {\bi v}={\bi 0},  \label{eq:2.23}
\en
where $p_1=p - \phi {\delta F}/{\delta\phi} 
+ f_{\rm FH} + {C}|\nabla \phi|^2/2$
 ensures the incompressibility condition (\ref{eq:2.18}). 
The  ${\bi {\cal F}}_{\rm p}$ is the force density 
acting on the polymer due to the fluctuation of $\phi$ and $\aw$ of the form, 
\be
{\bi {\cal F}}_{\rm p}  = 
-{\phi}{\nabla}{\frac{\delta F}{\delta\phi}} 
-{\frac{1}{4}}Q\nabla G +\nabla\cdot\asigma_{\rm p} .
\label{eq:2.25}
\en
The network stress tensor in (\ref{eq:2.22}) is expressed as    
\be
\sigma_{{\rm p} ij}=G\sum_k W_{ik}\delta W_{kj}+{\frac{1}{4}}GQ \delta_{ij} .
\label{eq:2.26}
\en

To express ${\bi v}_{\rm p}$ and ${\bi v}_{\rm s}$ 
in terms of $\phi$ and $\aw$, 
we assume slow processes and neglect the acceleration or inertial terms 
 in the two-fluid dynamic equations \cite{Doi-Onuki}. 
By setting $\p {\bi w}/\p t={\bi 0}$, we obtain 
\be
{\bi w}=\frac{1-\phi}{\zeta}{\bi {\cal F}}_{\rm p},\label{eq:2.27}
\en
where $\zeta$ is the friction coefficient 
between  polymer and  solvent and is 
estimated as $\zeta\sim 6\pi \eta_0 \xi_b^{-2}
\sim \eta_0 \phi^{2}/a^2 $  in the semidilute 
solution in terms of the blob size $\xi_b$.  
The mean  velocity ${\bi v}$ is expressed as 
\be 
{\bi v}= \av{\bi v}+ \bigg [ \frac{1}{-\eta_0\nabla^2} 
{\bi {\cal F}}_{\rm p}  \bigg ]_\perp, \label{eq:2.28}
\en 
where $\av{\bi v}$ is the mean  flow 
such as the shear flow in (\ref{eq:1.2}), 
$[\cdots]_\perp$ denotes taking the transverse part 
(whose Fourier component is perpendicular to the 
wave vector)  and the inverse operation 
$({-\eta_0\nabla^2})^{-1}$ may 
be expressed in terms of the  Oseen tensor 
in the limit of large system size \cite{Onukibook}.

In the following, we make our eqations dimensionless 
by measuring space and time in units of $\ell$ and $\tau_0$ defined by  
\be 
\ell= \frac{aN^{1/2}}{2 \sqrt{18}}, \quad 
\frac{1}{\tau_0}= \frac{4k_{\rm B} T }{\eta_0v_0 N^{3/2}},  
\label{eq:2.33}
\en 
where  $\ell$ is of the order of the gyration radius, 
and the time $\tau_0$ is the conformation 
relaxation time of a single chain in the dilute case. 
In our simulations 
the mesh size in numerical integration will be set equal to $\ell$. 
The velocities will be measured in units of $\ell/\tau_0$ and the 
stress components given in (\ref{eq:2.22}) will be  measured in units of 
\be 
\sigma_0 = k_{\rm B} T/(v_0 N^{3/2})= \eta_0/(4 \tau_0). 
\label{eq:2.34}
\en  
To avoid cumbersome notation, in the follwoing, 
we use  the same symbols for $t$, $\bi r$, $\nabla$, 
and the velocities even after rescaling.

\subsection{Rheological quantities} 

The  conformation tensor $\aw$ obeys (\ref{eq:2.19}) with $\tau^*$ 
being replaced by $\tau^*/\tau_0$ in the dimensionless form. 
Following Ref.\cite{Onukireview} we assume  
\be 
\tau^* / \tau_0 = (\Phi^4+ 0.2) /(1+Q),\label{eq:2.41}
\en 
where $Q=Q(\aw)$ is defined by (\ref{eq:2.13}) and 
the factor $1/(1+Q)$ accounts for quickening of the stress 
relaxation under large deformations \cite{commentR}.
Similarly, some authors assumed a deformation-dependent 
stress relaxation time in the rheological constitutive equations  
\cite{Larsonbook,Metzner}.  

Consequences of (\ref{eq:2.41}) are as follows. 
(i) In the dilute regime we have $\tau^* \cong 0.2 \tau_0$. 
(ii) The  relaxation time $\tau$ in (\ref{eq:2.21}) 
in the linear response regime becomes 
\be 
\tau =\tau_0 (\Phi^4+0.2). \label{eq:2.42}
\en 
(iii) The zero-frequency linear viscosity  
becomes 
\be 
\eta/\eta_0
=1+ \frac{1}{4}g \Phi^3 \tau/\tau_0 \cong  1+ \frac{1}{4}g\Phi^7, 
\label{eq:2.43}
\en 
where the small number $0.2$ in $\tau$ is omitted. 
(iv) Let us consider a homogeneous state under  shear, 
where ${\bi v}_{\rm p} ={\bi v} = \gdot y {\bi e}_x$ and $\phi=$const. 
In the high shear limit $\gdot\tau \gg 1$, 
by solving (\ref{eq:2.19}) we obtain shear thinning behavior, 
\be
\sigma_{{\rm p}xy} \sim g\Phi^3 (\gdot\tau)^{3/5}, 
\quad 
N_{{\rm p}1} \sim g\Phi^3 (\gdot\tau)^{4/5}.   \label{eq:2.46}
\en
Non-Newtonian behavior can arise from the factor $1/(1+Q)$ in $\tau^*$ 
even in homogeneous states.

Next we give dimensionless forms 
of the stress components for general inhomogeneous cases. 
From (\ref{eq:2.23}) and (\ref{eq:2.26}) 
the shear stress  $\sigma_{xy}$ 
and the normal stress difference $N_1= \sigma_{xx}- \sigma_{yy}$  
in units of $\sigma_0$  are written as 
\bea 
\sigma_{xy}&=& \sigma_{{\rm p}xy}- \frac{4}{\Phi}\nabla_x\Phi\nabla_y\Phi 
 +4 (\nabla_x v_y+ \nabla_y v_x),\hspace{5mm}\label{eq:2.47}
\\ 
N_1 &=& N_{{\rm p} 1}- \frac{4}{\Phi}
(\nabla_x\Phi\nabla_x\Phi - \nabla_y\Phi\nabla_y\Phi) \nonumber\\ 
&& +8 (\nabla_x v_x- \nabla_y v_y),\label{eq:2.48}
\ena 
where $\sigma_{{\rm p} xy}$ and $N_{{\rm p} 1}$ 
are the network  contributions, 
\bea 
\sigma_{{\rm p} xy} &=& g\Phi^3(W_{xx}+W_{yy}-1) W_{xy}, \label{eq:2.49}
\\ 
N_{{\rm p} 1} &=& g \Phi^3(W_{xx}+W_{yy}-1) (W_{xx}-W_{yy}).\label{eq:2.50}
\ena 
The second terms in (\ref{eq:2.47}) and (\ref{eq:2.48}) 
arise from inhomogeneity in $\Phi$ and give rise to the surface tension 
contributions in two-phase states \cite{O6}. 
The last terms are the usual viscous contributions.

The physical meaning of $\sigma_{{\rm p} xy}$ and $N_{{\rm p} 1}$ can be seen 
if they are related to the degree of chain extension. 
To this end let us decompose ${\delta}{\aw}$ as 
\be 
{\delta}{\aw} = w_1 {\bi e}_1 {\bi e}_1 + w_2 {\bi e}_2 {\bi e}_2 ,
\label{eq:2.51}
\en 
where $w_1$ and $w_2$ are the eigenvalues of $\delta{\aw}$ with $w_1 \ge w_2$, 
and 
\be 
{\bi e}_1= (\cos\theta,\sin\theta), \quad 
{\bi e}_2= (-\sin\theta,\cos\theta) \label{eq:2.52}
\en  
are the corresponding  eigenvectors with $\theta$ being the angle between  
the stretched direction and the $x$ axis.  
We may assume $-\pi/2 < \theta \le \pi/2$ without loss of generality.  
In terms of these quantities we obtain 
\bea 
\sigma_{{\rm p} xy} &=& \frac{g}{2}  \Phi^3 
(1+w_1 +w_2)  (w_{1}-w_{2})  \sin 2\theta , \label{eq:2.53}
\\ 
N_{{\rm p} 1} &=& g \Phi^3 
(1+w_1 +w_2) (w_{1}-w_{2})  \cos 2\theta. \label{eq:2.54}
\ena 
In weak shear $\tau \gdot \ll 1$, we have 
\be 
w_1-w_2 \sim \tau \gdot,\quad  \theta- \pi/4 \sim \tau \gdot,
\label{eq:2.55}
\en 
so $\sin 2 \theta \cong 1$ and $\cos 2\theta \sim \tau \gdot$, 
leading to the well-known results $\sigma_{{\rm p} xy} \sim \eta \gdot$ and 
$N_1 \sim \eta \tau \gdot^2$ in the Newtonian regime \cite{Larsonbook}. 
Here, for the analysis in Sec. III,  
we introduce an extension vector defined by 
\be 
{\bi w}({\bi r},t) = (w_1-w_2){\bi e}_1,  
\label{eq:3.1}
\en 
whose magnitude and direction represent 
the degree of chain extension and the extended direction.
In our simulations we shall see that the magnitudes of 
$w_1$ and $w_2$ are both considerably smaller than 1 at most space points. 
That is, the degree of extension is rather weak, but 
the network stress can overwhelm the viscous stress ($\propto \eta_0)$ 
because of the large factor $g\Phi^3$. 
This justifies the Gaussian form of $Q(\aw)$ in (\ref{eq:2.13}) in this work.

\section{Numerical Results }
\setcounter{equation}{0} 

We need numerical  approach to understand the nonlinear regime of 
shear-induced phase separation. 
To this end we integrate our model equations, 
(\ref{eq:2.14}) and (\ref{eq:2.19}), in the previous section 
on a $256\times 256$ lattice in two dimensions.  
The mesh size $\Delta x$ is set equal to $\ell$ 
in (\ref{eq:2.33}) \cite{spikes}. 
We use a numerical scheme developed 
by one of the present authors \cite{Onuki_comp,Ber},  
which uses the deformed coordinates $x'=x-\gdot t y$ and $y'=y$ and 
enables the FFT (fast Fourier transform) method 
to be carried out for shear flow. 
Here we impose the periodic boundary condition 
$f(x',y'+L)=f(x'+L,y')=f(x',y')$ for any quantity 
$f(x',y')$ in terms of the deformed coordinates.

As the initial condition at $t=0$ 
we assign Gaussian random numbers 
with variance (rms amplitude) 
$0.1$ to $\Phi$ at the lattice points. 
Then the initial $\Phi$ at each point consists of the average 
$\av{\Phi}$ and a random number. 
For $t>0$ we will solve the dynamic equations 
in the presence of shear 
without the random noise terms. 
Here, if quenching is is not very far above 
the coexistence curve,  
the initial variance is not very small 
and our choice 
can be appropriate \cite{initial_v}.

\subsection{Crossover from Newtonian to viscoelastic fluids 
below the coexistence curve} 

Even below the coexistence curve, if the shear rate is 
sufficiently strong ($\gdot \tau \gg 1$), 
the composition fluctuations vary in space gradually and there is no distinct  
phase separation. 
However, with increasing the quench depth (or $u$)  
and/or decreasing the shear rate $\gdot$, 
the shear-induced composition  fluctuations  
become composed of polymer-rich  and solvent-rich 
regions with sharp interfaces.  
Here we show that the domain morphology strongly 
depends on the shear modulus $G$.

In Fig.2 we show  phase separation patterns at $t=600$ 
and 2000 for $g=0$ (a), 0.01 (b), 0.1 (c), and 1 (d) 
after quenching at $t=0$, 
where $g$ represents the magnitude of $G$ in (\ref{eq:2.11}). 
The other parameters take common values as  $\av{\Phi}=2$, $u=3$, 
and $\gdot=0.025$. 
The initial Deborach number before phase separation is given by 
$\gdot \tau = \gdot \Phi^4 = 0.4$. 
In the phase separated semidilute regions we have  
\be 
\Phi_{\rm cx} \cong 3.70
\label{eq:3.3}
\en 
for $u=3$ on the polymer-rich branch of the coexistence curve 
in Fig.1, 
so within the polymer-rich domains the viscosity ratio in (\ref{eq:2.43}) 
is given by $1$ (a), $24.8$ (b), $239$ (c), and $2384$ (d).  
We can see gradual crossover from 
the patterns of the  Newtonian fluids to those of highly 
asymmetric viscoelastic fluids. 

In Fig.3 we show time evolution of the inverse of the 
perimeter length density for $g=$0, 0.01, 0.1, and 1, 
using the same conditions as in Fig.2. 
It may be regarded as the typical domain size $R(t)$. 
The domains continue to grow  for $g=0$ and $0.01$ 
within our  simulation time $t <10^4$,  
but the growth is gradually slowed down with increasing  $g$. 
For $g=1$ and $0.1$, we can see that the coarsening is nearly stopped.

\subsection{Domain size in two-phase flow}
Hereafter we fix $g$ at 1. 
In Fig.4 we show time evolution of the domain size $R(t)$ 
for $\av{\Phi}=2$ and  $u=3$ at various 
shear rates $\dot\gamma=$0.0005, 0.005, 0.025, and 0.05.
Remarkably, the coarsening is faster for $\dot\gamma=$0.005 
than for $\dot\gamma=$0.0005. 
This should be due to shear-induced coagulation of domains  
observed in near-critical fluids \cite{Onukireview,Baum2}, 
where shear accelerates collision and fusion of the domains. 
For shear rates larger than $0.025$, 
dynamical steady states are realized, 
where there should be a balance between the thermodynamic instability 
and shear-induced domain breakup as in the Newtonian case 
\cite{Onukibook,Onukireview}.

The characteristic domain size $R_{\rm D}$ 
in the steady states are of interest. 
In Newtonian immiscible mixtures 
under shear we have  $R_{\rm D} \sim C_{\rm New} 
/\av{\sigma_{xy}}$ in the low-Reynolds-number 
limit \cite{Onukibook,Onukireview,Hashimoto_ternary}, 
where $C_{\rm New}$ is of the order of the surface tension $\gamma$\cite{O7}.  
This formula  follows from a  balance 
between the surface energy density $\gamma/R_{\rm D}$ and the shear stress. 
Also in our  non-Newtonian case 
 Fig.5  suggests the same form,    
\be 
R_{\rm D} \sim C_{\rm vis} / \av{\sigma_{xy}}  
\label{eq:3.4}
\en 
at various $\gdot$. 
The coefficient  $C_{\rm vis} $ is independent of $\gdot$ 
but dependent on $\av{\Phi}$ as $20$, $13$, and $11$ 
for $\av{\Phi}=2.5$, 2, and 1.7, respectively. 
Fig.5 displays the product $\av{\sigma_{xy}}(t) R(t)$ 
as a function of time after quenching 
at  $g=1$ and  $u=3$ with   
$\langle\Phi\rangle=2$    
and  $\langle\Phi\rangle=2.5$.   
The curves  
tend to composition-dependent constants  
 independently of 
$\gdot$ at long times to confirm (\ref{eq:3.4}). 
However, we cannot derive (\ref{eq:3.4}) 
using simple arguments in this paper.

In  the viscoelastic case, as in (\ref{eq:2.47}) and (\ref{eq:2.48}), 
the stress consists of the three contributions. 
As shown in Fig.6 at $\gdot=0.05$, 
the network stress $\sigma_{{\rm p}xy}$ dominates 
over the gradient contribution in the polymer-rich regions, 
whereas the network, gradient, and viscous ones are of the same order 
in the interface regions. 
The particularly large size of $\sigma_{{\rm p}xy}$ 
in the polymer-rich regions suggests 
that these regions should behave like percolated gels 
and mostly support the applied stress. 
On the other hand, the viscous contribution is very small 
in the polymer-rich regions,  
but is nearly the  sole contribution in the solvent-rich regions.

\subsection{Fine domains at strong shear} 

For  the largest shear rate $\gdot=0.05$ in Fig.4, 
Fig.7 displays time evolution of the domains, 
where the polymer-rich domains are percolated  
and the angle of extension $\theta$ defined 
by (\ref{eq:2.51}) and (\ref{eq:2.52}) is close to $\pi/4$. 
These closely resemble the observed microscope pictures (in the $xz$ plane)
\cite{Moses,Hobbie}. 
The profile of $\Phi$ at the bottom of Fig.7 
shows that $\Phi$ becomes close to 0 in the solvent-rich regions, 
while it is around $\Phi_{\rm ex}$ in (\ref{eq:3.3}) 
in the polymer-rich regions. 

Fig.8 displays the structure factor $S(k_x,k_y) $ 
of the composition fluctuations in the steady state in Fig.7. 
It has sharp double peaks along the $k_x$ axis with peak wave number 
$k_{\rm p} \cong (2\pi/256)\times 9$, 
although we cannot see marked anisotropy in the shapes of domains. 
The origin of the peaks is that the domains are 
connected and hence are aligned perpendicularly to the flow direction 
on the spatial scale of $2\pi/k_{\rm p} \cong 28$.  

In addition, we notice that the profile of $\Phi$ 
(bottom one in Fig.7) 
exhibit spike-like behavior around some extrema,    
where $\nabla\Phi$ varies over distances of order $5$ \cite{spikes}.  
These steep changes  arise from convection due to the velocity fluctuations 
on such small scales in the nonlinear regime. 

Fig.9 shows the average stress components 
$\av{\sigma_{xy}}(t)$ and $\av{N_1}(t)$ vs time, 
which exhibit pronounced overshoots and subsequent noisy behavior. 
Here the network contributions 
in (\ref{eq:2.53}) and (\ref{eq:2.54}) are much larger  
than the surface tension contributions arising from the second 
terms in (\ref{eq:2.51}) and (\ref{eq:2.52}) 
by at least  one order of magnitude.   
This is consistent with Fig.6. 
Also shown is the average variance $\sqrt{\langle\delta\Phi^2\rangle}(t)$, 
which slowly increases over a long transient time of order  $2000$. 
This arises from  desorption of solvent  from the 
polymer-rich regions into the 
solvent-regions, as was observed by Tanaka and coworkers 
\cite{HTanaka93and94,HTanaka96}.   

\subsection{Large fluctuations  at weak shear} 

For the smaller shear  rate $\gdot=0.005$,  
Fig.10 demonstrates 
that the domain growth continues up to the system size 
in the simulation time $t=10^4$. 
Furthermore, comparing the two snapshots at $t=300$ and $500$,  
we can see that the domains are rotated as a whole in the early stage. 
Fig.11 shows that the chaotic temporal fluctuations 
of the stress are much more exaggerated than in Fig.9. 
Unusually the normal stress frequently takes negative values, 
while it is always positive at $\gdot=0.05$ as shown in Fig.9 
at any $\av{\Phi}$.

In Fig.12 
we further examine the origin of the strong fluctuations in this case. 
It displays the snapshots of the extension vector defined by (\ref{eq:3.1}), 
$\av{\sigma_{xy}}$, $\av{N_1}$, 
and the following rotationally invariant shear gradient \cite{S}, 
\be 
S=\bigg [ 2\sum_{ij}
 (\nabla_i v_j)^2 - (\nabla_xv_y-\nabla_y v_x)^2 \bigg ]^{1/2}. \label{eq:3.5}
\en 
Notice $S=0$ for pure rotation. 
At $t=500$, the angle $\theta$ exceeds  $\pi/4$ in most of 
the spatial points of the polymer-rich regions, 
resulting in  $\av{\cos 2\theta}=-0.21$ and $\langle N_1 \rangle=-0.44$. 
At $t=1500$, the points with $\theta < \pi/4$ constitute 
a majority in  the polymer-rich regions, 
leading to  $\langle N_1 \rangle=0.24$.
These  snapshots and those of the stress components 
clearly demonstrate the presence of {\it stress lines} forming networks,  
which are supported by the percolated polymer-rich regions and  
where the extension and the stress take large values.
The typical values of $w_1-w_2$ on these lines 
are $0.16$ at $t=500$ and $0.11$ at $t=1500$. 
For this shear rate we can see that $\langle N_1 \rangle$ becomes negative 
when the stress lines collectively rotate and the angle $\theta$ exceeds 
$\pi/4$ on these lines, while for $\gdot=0.05$ the stress lines are 
broken at  much faster rates and $\av{N_1}$ remains positive. 
On the other hand, $S$ in Fig.12 takes large values in the 
solvent-rich regions, indicating that the polymer-rich domains 
are largely rotated rather than being anisotropically deformed in shear.  
This tendency becomes conspicuous with decreasing shear. 
See the black regions of $S$ in Fig.12, where $\Phi$ is small and 
{\it slipping} is taking place.

Fig.13 shows a bird view of the velocity gradient $\p v_x/\p y$ 
(right) and the corresponding  snapshot of $\Phi$ at $t=2000$, 
where we choose $\gdot=0.025$, an intermediate shear rate in Fig.4. 
The other parameters are the same as in Figs.4-12.  
We recognize that the solvent-rich regions support this velocity gradient.
This means that 
strongly deformed solvent-rich domains act as a lubricant serving 
to diminish the macroscopic  stress or viscosity. 
This lubricant effect can be effective 
even for a small volume fraction of the solvent-rich domains 
because of the strong contrast of the viscoelastic properties 
in the two phases. 
In fact Wolf and Sezen \cite{Wolf_vis} presented 
this view to interpret their finding of a viscosity decrease which 
signals onset of phase separation in shear in semidilute solutions.  

\subsection{Domain morphology and rheology \\for various $\av{\Phi}$} 

Next we show that the domain morphology also strongly depends on the average 
polymer volume fraction $\av{\Phi}$.  
Fig.14 displays snapshots of $\Phi({\bf r},t)$ at  
$\gdot=0.05$, $u=3$, 
and $g=1$   for three compositions,
 $\langle \Phi\rangle=1.2$, 1.7, 2.0, and 2.5 (from left). 
The time is $t=1500$, where steady states are  
almost reached in all these cases.  
The profiles of $\Phi$ in the $x$ direction at $y=128$ are also shown 
in the upper part. We can see that $\Phi \cong \Phi_{\rm ex}\cong 3.7$ 
within the polymer-rich regions. With increasing $\av{\Phi}$, 
the collision frequency among the 
domains increases  and the domain size decreases. 
In fact, $R_{\rm D}$ 
determined from the perimeter length is 
 13.75, 10.83, and 6.67 for $\av{\Phi}=$1.7, 2.0, and 2.5, respectively. 
Even for $\langle \Phi\rangle=1.2$ the polymer-rich domains 
collide frequently and the domains shapes 
largely deviate  from sphericity. 
In Fig.15 we plot time evolutions of the average shear 
stress $\langle\sigma_{xy}\rangle$ for these values of $\langle \Phi\rangle$. 
In the dynamical steady state, the normalized  shear viscosity increase   
$\Delta\eta/\eta_0=\eta/\eta_0-1$ is given by 2, 4, 6.5 and 15 
for $\langle \Phi\rangle=1.2$, 1.7, 2.0 and 2.5, respectively.  
It is remarkable that the stress overshoot is nonexistent 
at small $\av{\Phi}$ and gradually develops with increasing $\av{\Phi}$.

\subsection{Deformation of sponge-like domains at small shear}

Finally  we examine how the sponge-like domain structure observed by 
Tanaka and coworkers \cite{HTanaka93and94,HTanaka96} is deformed by shear flow.
On the left of Fig.16 we show one example 
of such a domain structure without shear for $\langle\Phi\rangle=1.2$, 
$u=3$,  and $g=1$, where the polymer-rich regions 
are percolated despite  relatively small $\av{\Phi}$ 
\cite{Taniguchi,Tanaka-Araki}. 
The right  part is the result under shear 
$\gdot=0.005$ with the other parameters being common, 
where the applied strain $\gdot t$ is $1, 2, 5,$ and 10 
at $t=200$, 400, 1000, and 2000 for the given snapshots, respectively. 
For this weak shear rate,  
the initial stage of phase separation is not much different 
from the case without shear but the coarsening is quickened  
as in Fig.4 for $\av{\Phi}=2$. 
Here the polymer-rich domains are gradually thickened 
but remain highly extended in the flow direction, 
while the domains in Fig.14(a) are torn into pieces because of 
larger shear $\gdot=0.05$. 
Also in this case the  average shear stress $\av{\sigma_{xy}}(t)$ 
and normal stress difference $\av{N_1(t)}$ exhibit chaotic behavior  
and the product $\av{\sigma_{xy}}(t) R(t)$ tends to a constant ($\sim 2.4)$. 
Again as in Fig.11, the negativity of the normal stress difference 
is conspicuous in the early stage (not shown here).

\section{Summary and concluding remarks}

Though performed in two dimensions, 
we have numerically solved the two-fluid dynamic model 
of sheared semidilute polymer solutions with theta solvent, 
where the chain deformations are represented by 
the conformation tensor $ W_{ij}$.  
The free energy density 
depends on $ W_{ij}$ as well as the polymer volume fraction $\phi$ 
as in (\ref{eq:2.8}).
In our simulations, 
the initial value of the composition is 
random, 
but the thermal noise terms in the dynamic equations 
are neglected for $t>0$. 
We  summarize our main results.\\  
(i) With varying the parameter 
$g$ representing the magnitude of the shear modus, 
we have examined spinodal decomposition as in Figs.2 and 3, 
which show the crossover of the domain growth 
from Newtonian to viscoelastic fluids. 
The  domain growth  is nearly  stopped 
for $g \gs 0.1$ at relatively large shear within our simulation time.\\ 
(ii) The domain growth in spinodal decomposition 
has been examined at $g=1$ with varying $\gdot$ in Fig.4. 
The domain size $R_{\rm D}$ in steady states becomes finer with 
increasing $\gdot$ and, as demonstrated in Fig.5, 
it satisfies the relation (\ref{eq:3.4}). 
Its  theoretical derivation is not given in this paper.   
In Fig.6 the network, gradient, and viscous stress 
contributions are compared, 
which indicates 
that the polymer-rich regions mostly support the applied stress.  
In Fig.9 the variance $\sqrt{\langle\delta\Phi^2\rangle}(t)$ 
continues to increase even after the saturation of the stress 
and the domain size, which arises from slow desorption of solvent 
from the polymer-rich regions \cite{HTanaka93and94,HTanaka96}. 
\\
(iii) Fig.7 displays the composition patterns 
at relatively large shear $\gdot=0.05$. 
Its structure factor in Fig.8 suggests that 
the domains are more correlated 
along the shear-gradient direction 
rather than along the flow direction. This is consistent with the experiments 
\cite{Wu,Saito,Moses,Hobbie}.\\ 
(iv) At  $\gdot=0.05$, 
$\Phi$ fluctuates in the region $|\Phi-\av{\Phi}|$ 
between $0$ and $\Phi_{\rm ex}$ in Fig.7, 
 exhibiting the spike-like behavior. 
Another notable difference is 
that the variance saturates  rapidly above the coexistence curve 
\cite{Onukireview,Onuki97} 
but slowly at a larger value in Fig.9.\\  
(v) At much smaller shear  $\gdot=0.005$,  
the domain size becomes larger as in Fig.10 
and the stress fluctuations 
look much enlarged as in Fig.11. 
as compared to the stress curves in Fig.9 for $\gdot=0.05$. 
For this small shear, 
percolated stress lines are formed in the polymer-rich regions 
as in Fig.12. 
There, we also notice that the non-rotational velocity gradient 
$S$ defined by (\ref{eq:3.5}) becomes large in the solvent-rich regions. 
Fig.13 shows that 
the velocity gradient $\p v_x/\p y$ takes large values 
in the solvent-rich regions, suggesting that they act as a lubricant.\\
(vi) Fig.14 illustrates how the domain morphology depends on the  
average polymer volume fraction $\av{\Phi}$. Fig.15 
shows that the stress-strain curves also strongly depends on  $\av{\Phi}$.\\ 
(vii) Fig.16 shows time evolution of the 
sponge-like patterns without shear \cite{HTanaka93and94,HTanaka96} 
and under weak shear.\\ 

Furthermore, we make comments on experimental aspects.\\   
(i) We have calculated the space averages of the stress components. 
In usual rheology experiments, however, 
the stress acting on the boundary surface is measured. 
In  a dynamical steady state, 
 the time averages  of the 
space-averaged stress 
 and the surface stress  do coincide, but their  
 time-dependent fluctuations  
can be different.  We note that the  
 time-dependent  fluctuations 
in these quantities  become  significant 
when the fluctuating entities are of very large sizes. 
Here it is worth noting that 
more than three decades ago Lodge \cite{Lodge} reported
abnormal temporal fluctuations of the normal 
stress difference at a hole of 1 mm diameter from a polymer solution contained 
in a cone-plate apparatus.
He ascribed its origin to growth of inhomogeneities or gel-like particles.\\  
(ii) Below the coexistence curve, 
we have shown sharp stress overshoots at relatively strong shear 
in Fig.9.  
However, the stress-strain curves at relatively small shear in Fig.11 
and those corresponding to Fig.16 (not shown) are more complex, 
where the frequent negativity of the normal stress is conspicuous.  
In the  snapshots  in  Figs. 7, 10, and 16 
the domains are collectively rotated as a whole in the early stage. 
Thus the first normal stress  might become negative (at least) 
after its first peak.\\ 
(iii) Disappearance of the stress peaks 
at small average volume fractions in Fig.15 should also be observed.\\
(iv) Experiments are needed on establishment of dynamical steady states 
with the characteristic heterogeneity length given by (\ref{eq:3.4}).  
Strong deviations of the polymer-rich domains from sphericity 
are also characteristic. 
These results  gained from the simulations could be compared with 
microscope observations \cite{StringPRL,Moses}.  
\\ 
In future, we should perform simulations to examine heterogeneous 
fluctuations in various sheared complex fluids such as polymer blends 
or wormlike micellar systems. 

\section*{Acknowledgments}
We would like to thank Prof. Takeji Hashimoto and 
Prof. Hajime Tanaka for valuable discussions. 
This work is supported by 
Grants in Aid for Scientific 
Research 
and for the 21st Century COE project 
(Center for Diversity and Universality in Physics)
 from the Ministry of Education, 
Culture, Sports, Science and Technology of Japan.


\widetext
\newpage
\begin{figure}[h]
\includegraphics[width=0.6\linewidth]{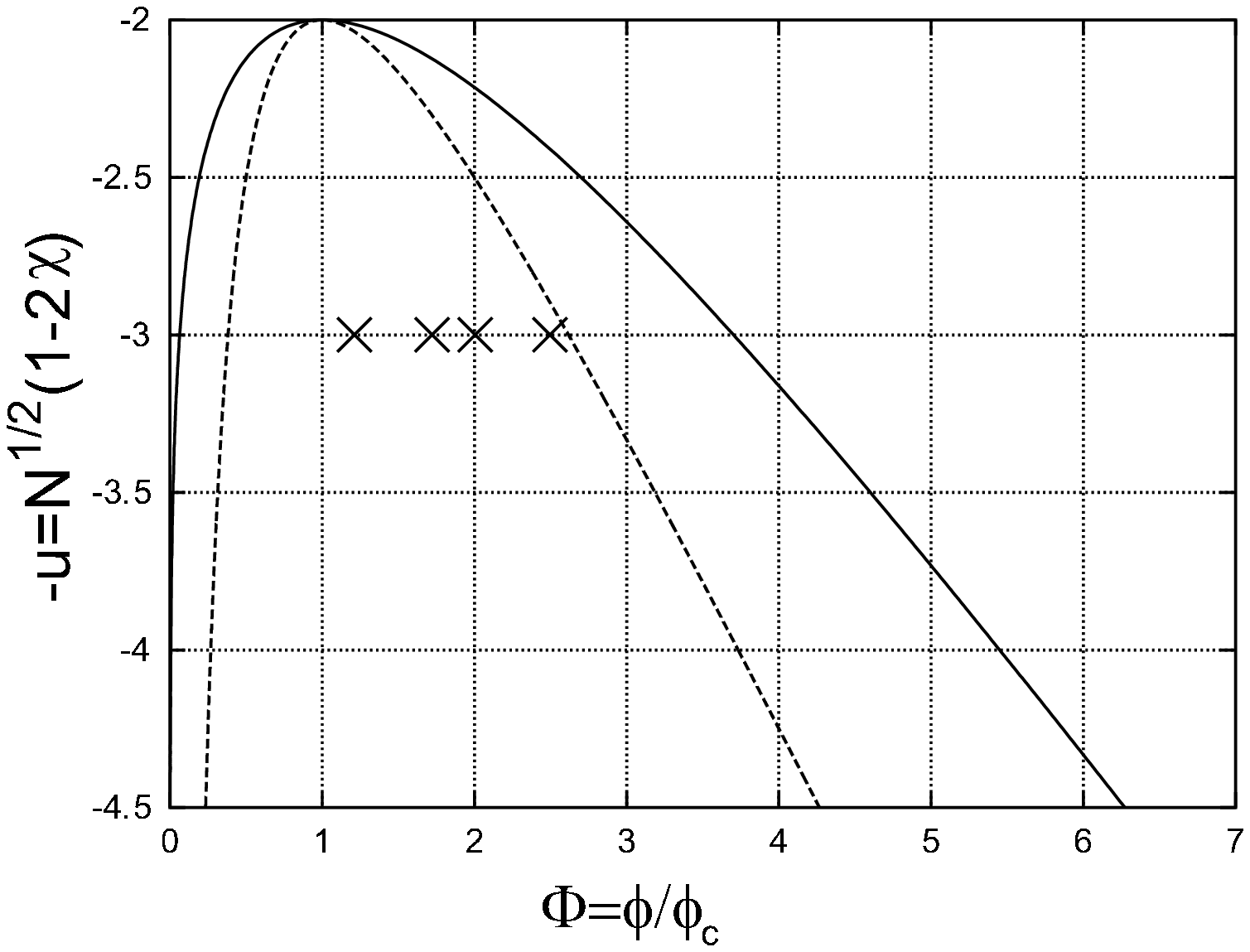}
\caption{\protect
\small{
Coexistence curve (solid line) and  spinodal curve (dashed line) 
for polymer solutions obtained from (\ref{eq:2.1})
 in the plane of $-u=N^{1/2}(1-2\chi)$ and $\Phi=\phi/\phi_c$.
The points ($\times$) represent the initial states 
of our simulations.
}}
\label{Fig1}
\end{figure}

\begin{figure}[h]
\includegraphics[width=0.7\linewidth]{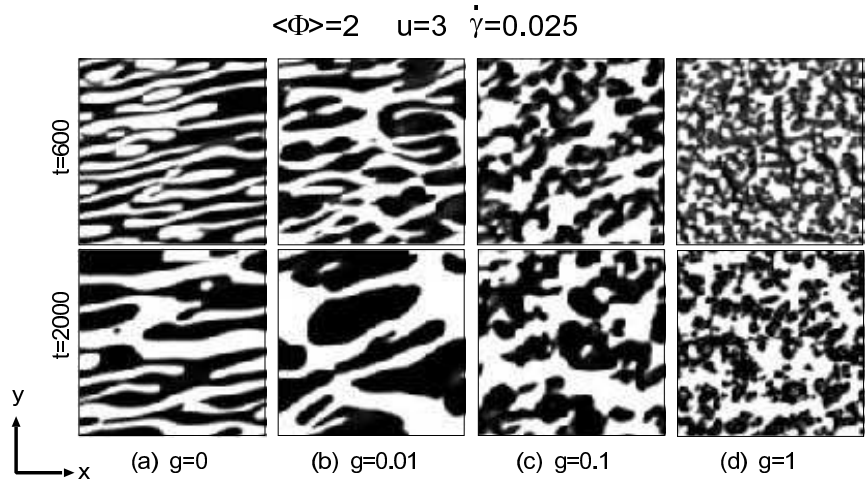}
\caption{\protect
\small{
Crossover of domain  patterns of $\Phi({\bf r},t)$ below the spinodal curve 
with increasing $g$ from 0 to 1 for $u=3$ and  $\langle\Phi\rangle=2$   
in shear flow with   $\dot\gamma=0.025$.
}}
\label{Fig2}
\end{figure}

\begin{figure}[h]
\includegraphics[width=0.5\linewidth]{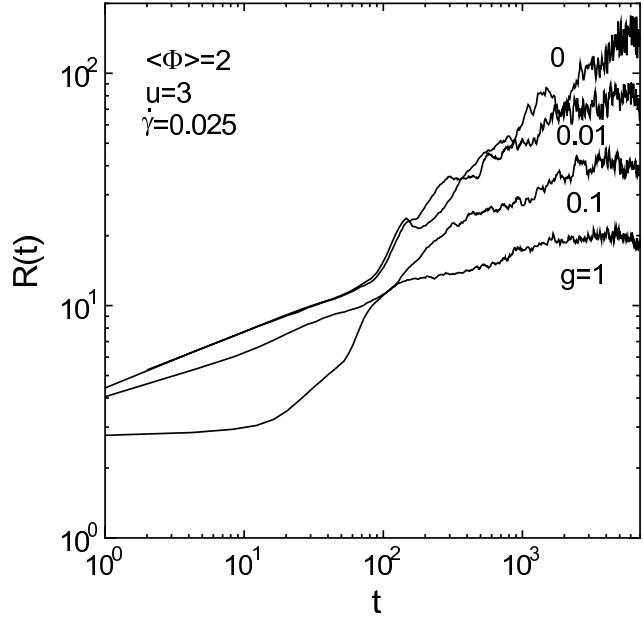}
\caption{\protect
\small{
Time evolution of the domain size $R(t)$ 
($=$ the inverse of the perimeter length density) 
below the spinodal curve with increasing the shear modulus 
as $g=$0, 0.01, 0.1, and 1 in shear flow with $\dot\gamma=0.025$. 
The other parameters are the same as those in Fig.2. 
The domain growth is nearly stopped 
for $g=1$ and $0.1$ within the simulation time.
}}
\label{Fig3}
\end{figure}

\begin{figure}[h]
\includegraphics[width=0.5\linewidth]{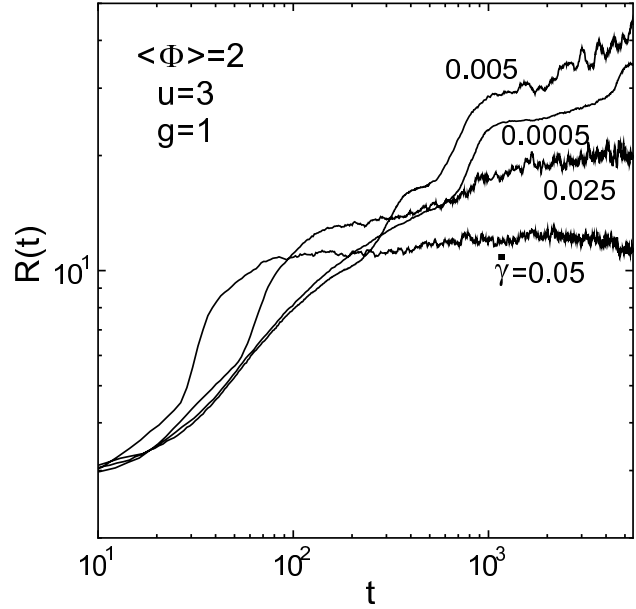}
\caption{\protect
\small{
Time evolution of the domain size $R(t)$ 
below the spinodal curve for  $g=1$, $u=3$, 
and $\langle\Phi\rangle=2$. 
Here $\dot\gamma=$0.0005, 0.005, 0.025, and 0.05.  
At small shear rates flow-induced coagulation 
accelerates the domain growth as demonstrated by the curve of $\gdot=0.005$. 
For $\gdot \gg 0.005$, shear-induced 
domain breakup becomes dominant and dynamical steady states  
are realized at smaller domain sizes.
}}
\label{Fig4}
\end{figure}

\begin{figure}[h]
\includegraphics[width=0.6\linewidth]{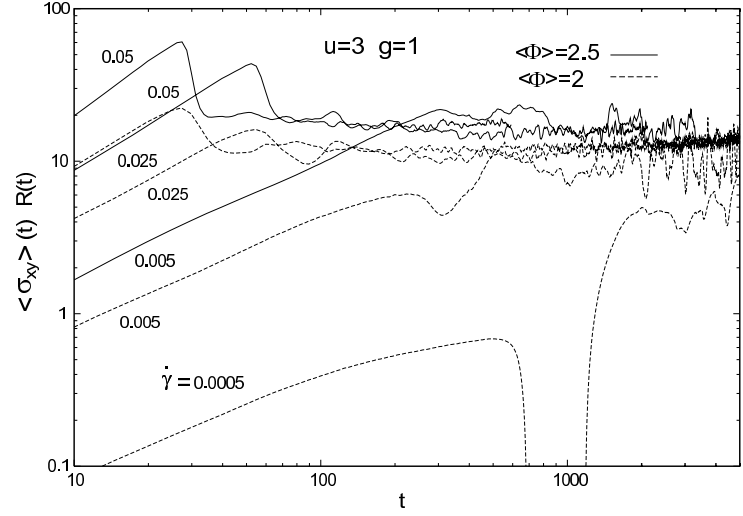}
\caption{\protect
\small{
Time evolution of  $\av{\sigma_{xy}}(t) R(t) $ 
below the spinodal curve  at various shear rates  
for $g=1$ and  $u=3$. Here  $\langle\Phi\rangle=2$ (solid lines) 
or $\langle\Phi\rangle=2.5$ (dotted  lines).   
At long times the curves tend to 
fluctuate around $15-20$  confirming (3.4).  
}}
\label{Fig5}
\end{figure}

\begin{figure}[h]
\includegraphics[width=0.6\linewidth]{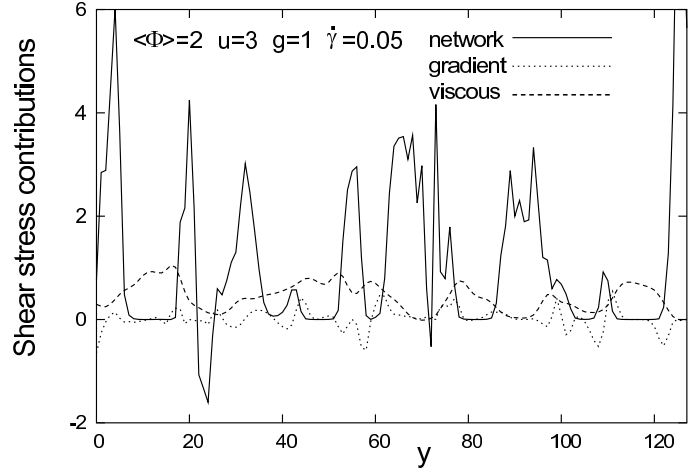}
\caption{\protect
\small{
Three contributions, $\sigma_{{\rm p}xy}$ (network), 
$-4(\nabla_x\Phi)(\nabla_y\Phi)/\Phi$ (gradient), 
and $4(\nabla_xv_y+\nabla_yv_x)$ (viscous) 
to the local shear stress  $\sigma_{xy}(x,y)$ 
in (2.40) in the range $0<y<128$ at $x=0$ 
for $u=3$, $\langle\Phi\rangle=2$, and $\gdot=0.05$ .  
The network stress is overwhelming in the polymer-rich regions, 
while the viscous one is relatively large in the solvent-rich regions.   
The gradient contribution is appreciable only 
in the interface regions, where the three contributions are of the same order. 
}}
\label{Fig6}
\end{figure}

\begin{figure}[h]
\includegraphics[width=0.4\linewidth]{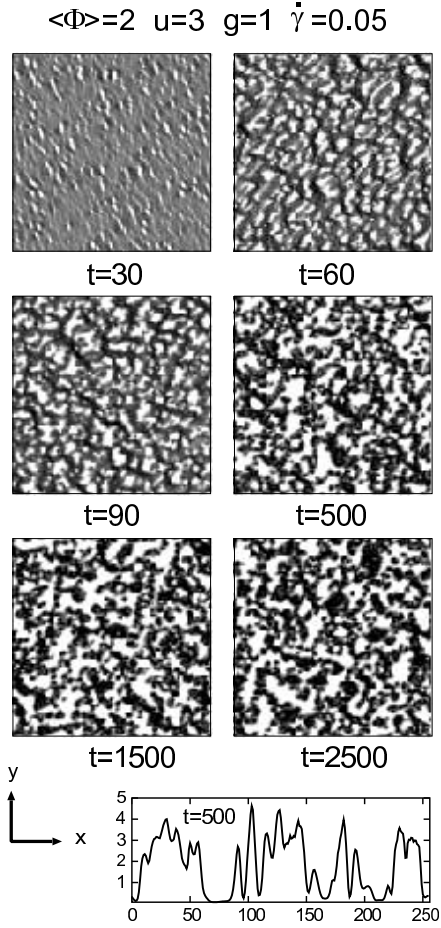}
\caption{\protect
\small{
Time evolution of  $\Phi({\bf r},t)$ 
at $\dot\gamma=0.05$ for $u=3$, $\langle\Phi\rangle=2$, 
and $g=1$ below the spinodal curve.  
The bottom figure is the profile at $t=500$ in the $x$ 
direction at $y=128$. 
The domain size evolution is given by the curve of the largest shear 
$\gdot=0.05$ in Fig.4.
}}
\label{Fig7}
\end{figure}

\begin{figure}[h]
\includegraphics[width=0.7\linewidth]{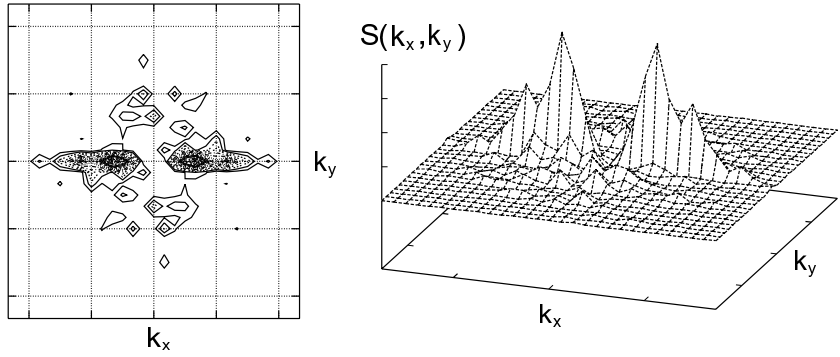}
\caption{\protect
\small{
Structure factor $S({\bi k})$ for the composition patterns 
 in Fig.7 in the dynamical steady state. 
}}
\label{structurefactor}
\end{figure}

\begin{figure}[h]
\includegraphics[width=0.55\linewidth]{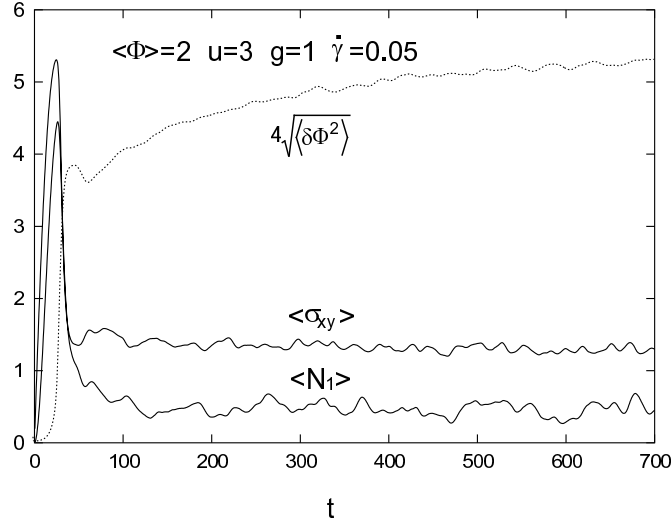}
\caption{\protect
\small{
Time evolution of the average shear stress $\langle\sigma_{xy}\rangle(t)$,  
the average normal stress difference $\langle N_1\rangle(t)$, 
and the average variance $\sqrt{\langle\delta\Phi^2\rangle}(t)$ (dotted line)  
at $\dot\gamma=0.05$ for $u=3$, $\langle\Phi\rangle=2$, and $g=1$ 
below the spinodal curve.
}}
\label{Fig9}
\end{figure}

\begin{figure}[h]
\includegraphics[width=0.4\linewidth]{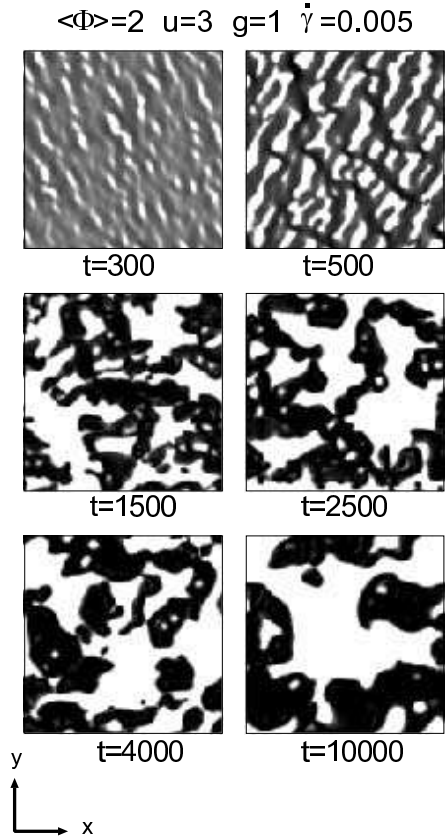}
\caption{\protect
\small{
Time evolution of $\Phi({\bf r},t)$ at $\dot\gamma=0.005$  for 
$u=3$, $\langle\Phi\rangle=2$, and $g=1$ 
below the spinodal curve.  For this 
 shear rate Fig.4  shows that 
the domain size increases up to the system size at $t=10^4$. 
}}
\label{Fig10}
\end{figure}

\begin{figure}[h]
\includegraphics[width=0.6\linewidth]{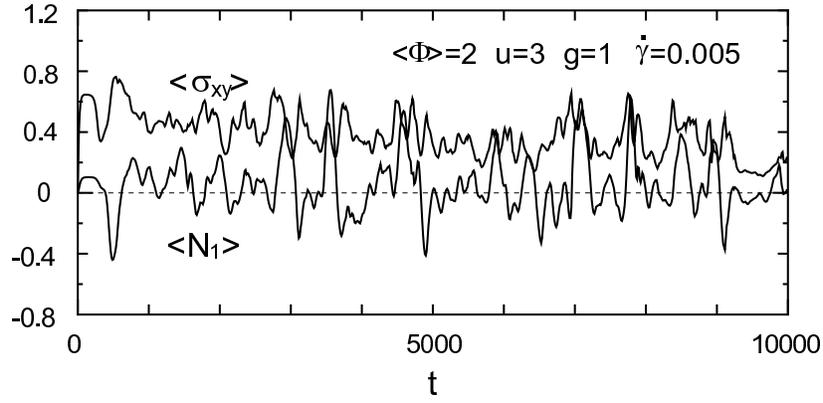}
\caption{\protect
\small{
Chaotic time evolution of 
the average shear stress $\langle\sigma_{xy}\rangle(t)$ and 
the average normal stress difference $\langle N_1 \rangle(t)$ 
at $\dot\gamma=0.005$ in the run which produced  Fig.10. 
For this weak  shear the deviatoric stress components  
 exhibit large fluctuations and $\langle N_1 \rangle(t)$ frequently 
takes  negative values.
}}
\label{Fig11}
\end{figure}

\begin{figure}[h]
\includegraphics[width=0.7\linewidth]{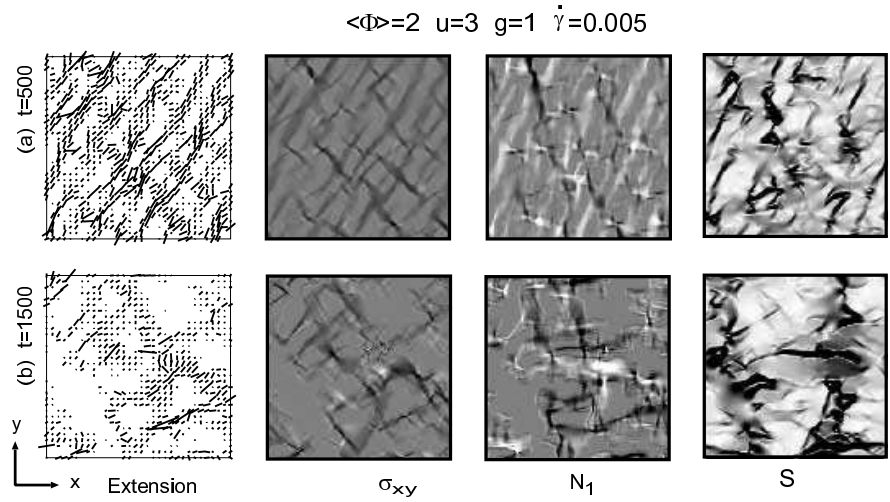}
\caption{\protect
\small{
Patterns of the extension, the shear stress, the normal stress difference, 
and the rotationally  invariant shear gradient (\ref{eq:3.5})  
at $t=500$ (top) and $t=1500$ (bottom). The corresponding 
composition  patterns are shown in Fig.10. 
We can see {\it stress lines} with large values of the extension and 
the deviatoric stress components.  The $S$ is large 
in solvent-rich slipping regions.
}}
\label{Fig12}
\end{figure}

\begin{figure}[h]
\includegraphics[width=0.54\linewidth]{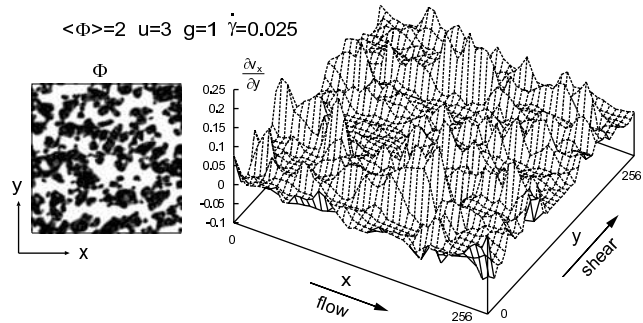}
\caption{\protect
\small{
$\Phi({\bf r},t)$ (left) and $\partial v_x({\bf r},t)/\partial y$ (right) 
at $\dot \gamma=0.025$ for  $u=3$, $\langle\Phi\rangle=2$, and $g=1$ 
in a steady state. 
This velocity gradient takes large values where slipping is taking place 
in the percolated solvent-rich region.
}}
\label{Fig13}
\end{figure}

\begin{figure}[h]
\includegraphics[width=0.62\linewidth]{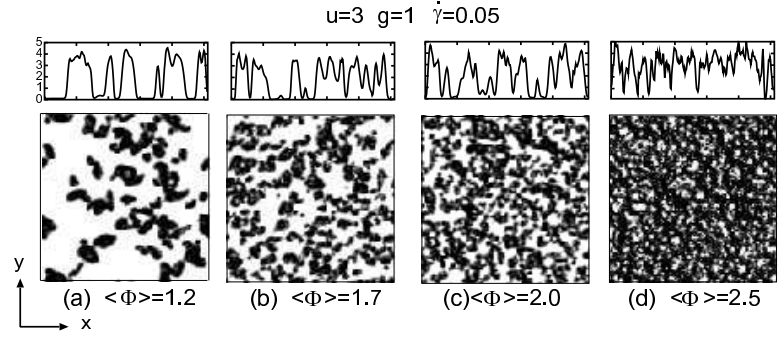}
\caption{\protect
\small{
Snapshots of $\Phi({\bf r},t)$ 
for $\langle \Phi\rangle=1.2$, 1.7, 2.0, and 2.5 
in steady states at $\gdot=0.05$.  
The profiles of $\Phi$ in the $x$ direction at $y=128$ are also shown.
}}
\label{Fig14}
\end{figure}

\begin{figure}[h]
\includegraphics[width=0.5\linewidth]{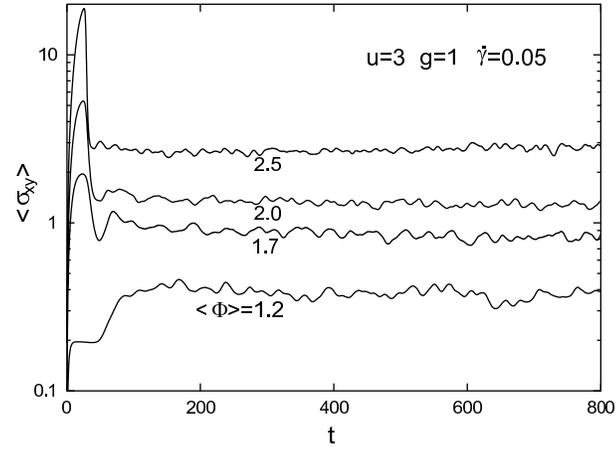}
\caption{\protect
\small{
Time evolution of $\langle\sigma_{xy}\rangle (t)$
 at various $\langle\Phi\rangle$.  
The corresponding composition patters are shown in Fig.14. 
The overshoot disappears at small $\av{\Phi}$. 
}}
\label{Fig15}
\end{figure}

\begin{figure}[h]
\includegraphics[width=0.63\linewidth]{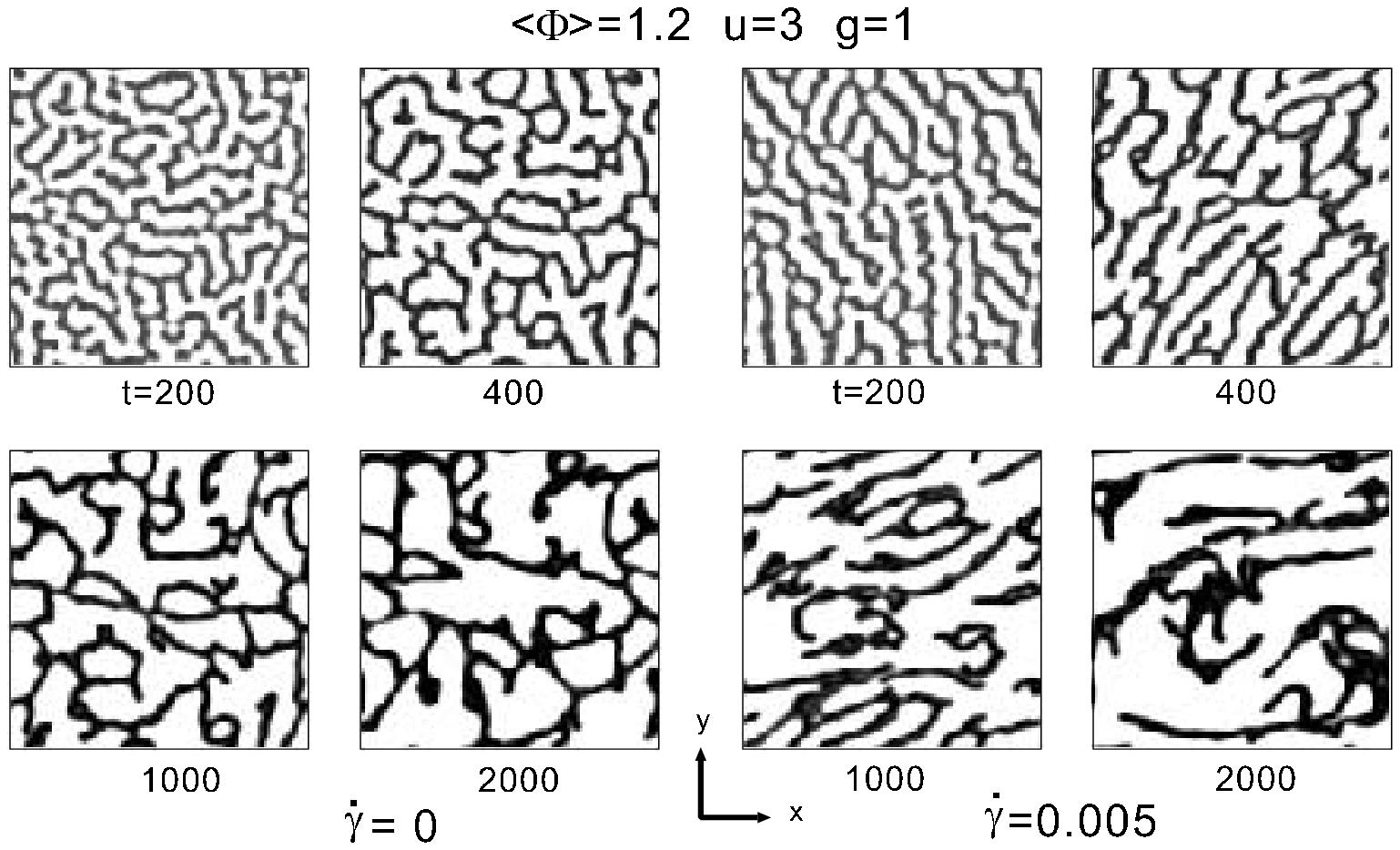}
\caption{\protect
\small{
Sponge-like domain structures without shear (left) 
and under shear $\gdot=0.005$ (right) for $\langle\Phi\rangle=1.2$, 
$u=3$, and $g=1$. 
}}
\label{Fig16}
\end{figure}


\begin{references}
\bibitem{spikes} 
We have also performed simulations with 
the mesh size taken to be $\Delta x= \ell/\sqrt{2}$ and $\ell/2$. 
Then the curves  in the bottom plate in 
Fig.7 coincide within 5$\%$ 
without any qualitative differences. 
The behavior of $\Phi$ in the spike-like regions  
in Fig.7 are well described even with $\Delta x= \ell$.
\bibitem{initial_v} At $t=0$ 
we consider  the variance of 
the deviation $\int d{\bi r}\delta \phi/\phi_{\rm c}\ell^d$ 
averaged in one unit cell. 
Its thermal value  is of order $\Phi^{-1/2}$ 
in the semidilue theta condition.
\bibitem{S} 
Under (\ref{eq:2.18}), $S$ in (\ref{eq:3.5}) and $\nabla_xv_y-\nabla_y v_x$ 
are the strain rates invariant with respect to 
 rotation of the reference frame in two dimensions.
\end{references}
\end{document}